\pdfoutput=1
\documentclass[aip,reprint,apl]{revtex4-1}
\usepackage{graphicx}
\usepackage{amsmath}
\usepackage{amssymb}
\usepackage{booktabs}
\usepackage{array}
\usepackage{xcolor}
\usepackage{acronym}
\usepackage{pifont}
\usepackage[normalem]{ulem}

\begin{document}


\title{Electron effective mass in Sn-doped monoclinic single crystal $\beta$-gallium oxide determined by mid-infrared optical Hall effect} 



\author{Sean~Knight} 
\email[Electronic mail: ]{sean.knight@engr.unl.edu}
\affiliation{Department of Electrical and Computer Engineering, University of Nebraska-Lincoln, Lincoln, NE 68588, USA}
\author{Alyssa~Mock}
\affiliation{Department of Electrical and Computer Engineering, University of Nebraska-Lincoln, Lincoln, NE 68588, USA}
\author{Rafa\l{}~Korlacki}
\affiliation{Department of Electrical and Computer Engineering, University of Nebraska-Lincoln, Lincoln, NE 68588, USA}
\author{Vanya~Darakchieva}
\affiliation{Terahertz Materials Analysis Center, Department of Physics, Chemistry and Biology (IFM), Link\"{o}ping University, SE 58183, Link{\"o}ping, Sweden}
\author{Bo~Monemar}
\affiliation{Terahertz Materials Analysis Center, Department of Physics, Chemistry and Biology (IFM), Link\"{o}ping University, SE 58183, Link{\"o}ping, Sweden}
\affiliation{Institute of Global Innovation Research, Tokyo University of Agriculture and Technology, Koganei, Tokyo, Japan}
\author{Yoshinao~Kumagai}
\affiliation{Institute of Global Innovation Research, Tokyo University of Agriculture and Technology, Koganei, Tokyo, Japan}
\affiliation{Department of Applied Chemistry, Tokyo University of Agriculture and Technology, Koganei, Tokyo, Japan}
\author{Ken~Goto}
\affiliation{Department of Applied Chemistry, Tokyo University of Agriculture and Technology, Koganei, Tokyo, Japan}
\affiliation{Tamura Corporation, Sayama, Saitama, Japan}
\author{Masataka~Higashiwaki}
\affiliation{National Institute of Information and Communications Technology, koganei, Tokyo, Japan}
\author{Mathias~Schubert}
\affiliation{Department of Electrical and Computer Engineering, University of Nebraska-Lincoln, Lincoln, NE 68588, USA}
\affiliation{Terahertz Materials Analysis Center, Department of Physics, Chemistry and Biology (IFM), Link\"{o}ping University, SE 58183, Link{\"o}ping, Sweden}
\affiliation{Leibniz Institute for Polymer Research, Dresden, Germany}



\date{\today}

\begin{abstract}
The isotropic average conduction band minimum electron effective mass in Sn-doped monoclinic single crystal $\beta$-Ga$_2$O$_3$ is experimentally determined by mid-infrared optical Hall effect to be $(0.284\pm0.013)m_{0}$ combining investigations on (010) and ($\bar{2}01$) surface cuts. This result falls within the broad range of values predicted by theoretical calculations for undoped $\beta$-Ga$_2$O$_3$. The result is also comparable to recent density functional calculations using the Gaussian-attenuation-Perdue-Burke-Ernzerhof hybrid density functional, which predict an average effective mass of $0.267m_{0}$ (arXiv:1704.06711 [cond-mat.mtrl-sci]). Within our uncertainty limits we detect no anisotropy for the electron effective mass, which is consistent with most previous theoretical calculations. We discuss upper limits for possible anisotropy of the electron effective mass parameter from our experimental uncertainty limits, and we compare our findings with recent theoretical results. 
\end{abstract}

\pacs{}
\maketitle 

Single crystal gallium (III) oxide is a desirable material for optical and electronic applications due to its unique physical properties such as its transparent conducting nature and wide band gap.\cite{StepanovRAMS44_2016} As a transparent conductor, Ga$_2$O$_3$ is useful for various types of transparent electrodes, for example in flat panel displays,\cite{Betz_2006} smart windows,\cite{Granqvist_1995,Gogova_1999} photovoltaic cells,\cite{Granqvist_1995} and gas sensors.\cite{Reti_1994} Due to its wide band gap, Ga$_2$O$_3$ has a larger breakdown voltage than SiC and GaN, which makes it an excellent candidate for power devices.\cite{HigashiwakiSST31_2016,Wager_2003,Sturm_2016,Sturm_2015,Furthmuller_2016} Among the five phases, the monoclinic $\beta$ phase is the most stable, and is expected to possess highly anisotropic properties which may prove useful for various applications.\cite{Roy_1952,Tippins_1965} 

Precise knowledge of the free charge carrier properties is imperative for electronic and optoelectronic device design and operation. Experimentally determined results for effective mass, free charge carrier concentration, and mobility parameters are currently scarce for $\beta$-Ga$_2$O$_3$. Numerous theoretical investigations have yielded a wide range of values for the electron effective mass: from $0.12m_{0}$ to $0.39m_{0}$, where $m_\textrm{0}$ is the free electron mass.\cite{he2006,yamaguchi2004,MockarXivGa2O3_2017,Furthmuller_2016,varley2010,Peelaerspssb2015Ga2O3meff,he2006_2,Ming-GangJMCA2014} Most calculations predict only minimal anisotropy. Although, recent Gaussian-attenuation-Perdue-Burke-Ernzerhof (Gau-PBE) hybrid density functional calculations predict slightly higher anisotropy.\cite{MockarXivGa2O3_2017} Using a combination of optical transmission and electrical Hall effect measurements, the authors in Ref.~\onlinecite{UedaAPL71_1997} estimate a range of values for electron effective mass parameter along the \textbf{b} and \textbf{c} crystal directions to be $m^{\ast}_{\textrm{b}}=0.5m_{0}~\textrm{to}~1.0m_{0}$ and $m^{\ast}_{\textrm{c}}=1.0m_{0}~\textrm{to}~2.0m_{0}$, respectively. Electrical Hall effect measurements on $\beta$-Ga$_2$O$_3$ allow access to free charge carrier concentration and mobility,\cite{SuzukiPSSC4_2007,IrmscherJAP110_2011,VilloraAPL92_2008,UedaAPL71_1997} but this technique alone cannot resolve the effective mass parameter.

The optical Hall effect is a physical phenomenon exploited in our  measurement technique, which employs generalized spectroscopic ellipsometry in combination with external magnetic fields to obtain the free charge carrier properties of semiconducting materials without electrical contacts.\cite{Philipp_RSI,Schubert:03,hofmann_pssc03,hofmann_jem08,Schubert_OHE} This technique measures the change in the polarization of light after interaction with a sample due to a Lorentz force acting on the free charge carriers. In contrast with the electrical Hall effect, the optical Hall effect is capable of obtaining the effective mass, carrier concentration, mobility, and charge carrier type parameters simultaneously.

In this work, we experimentally determine the electron effective mass in Sn-doped monoclinic single crystal $\beta$-Ga$_2$O$_3$ by mid-infrared optical Hall effect (MIR-OHE) measurements. To our best knowledge, determination of free charge carrier parameters in crystals with monoclinic symmetry by magneto-optical methods has not been reported yet. We compare our results to values reported in previous theoretical and experimental work, and we discuss the anisotropy of the effective electron mass parameter. Here we find no discernible anisotropy and assume an isotropic average parameter. We discuss the amount of finite anisotropy that may remain hidden within our present experimental error bars for the effective mass to be potentially discovered by subsequent experiments.

Two surface cuts, (010) and ($\bar{2}01$), of Sn-doped single crystal $\beta$-Ga$_2$O$_3$ are investigated in this work. The crystals were grown using the edge-defined film-fed growth method by Tamura Corp. (Japan).\cite{Aida_2008,Sasaki_2012,Shimamura_2013} The dimensions for the (010) surface are (0.65$\times$10$\times$10)~mm, and (0.65$\times$10$\times$15)~mm for the ($\bar{2}01$) surface. The optical response of $\beta$-Ga$_2$O$_3$ is governed by the monoclinic Cartesian dielectric function tensor.\cite{Schubert_2016} Here the Cartesian direction \textbf{x} is contained within the sample surface plane and is oriented along the propagation direction of light incident on the sample. The \textbf{z} direction is oriented into the sample surface. The crystal directions in $\beta$-Ga$_2$O$_3$ are denoted \textbf{a}, \textbf{b}, and \textbf{c}, where the monoclinic angle $\beta$ = 103.7$^{\circ}$ lies between \textbf{a} and \textbf{c}.\cite{Geller_1960} We choose to align \textbf{a} and \textbf{b} along \textbf{x} and \textbf{-z}, respectively, such that \textbf{c} lies within the \textbf{x-y} plane. For practicality, we introduce the direction \textbf{c$^{\star}$} parallel to \textbf{y}, so that \textbf{a}, \textbf{b}, and \textbf{c$^{\star}$} form a pseudo-orthorhombic system. We define azimuth angle $\phi$ as a rotation about the \textbf{z} axis for a given crystal axes orientation.\cite{Schubert_2016} For the (010) surface, $\phi$~=~0$^{\circ}$ corresponds to \textbf{a} aligned along \textbf{x}. For the ($\bar{2}01$) surface, $\phi$~=~0$^{\circ}$ corresponds to \textbf{b} aligned along \textbf{y}.

Generalized spectroscopic ellipsometry is the measurement technique employed here to determine the free charge carrier properties of $\beta$-Ga$_2$O$_3$. Ellipsometric data is obtained using the Mueller matrix formalism.\cite{Fujiwara_2007,schubert_prb96} WVASE (J.A. Woollam Co. Inc.) is used to acquire and analyze the data. The MIR-OHE data is measured using a home-built Fourier transform infrared ellipsometer in the spectral range of 550~cm$^{-1}$ to 1500~cm$^{-1}$ with a resolution of 2~cm$^{-1}$. The home-built ellipsometer is capable of attaining the upper-left 3~$\times$~3 block of the complete 4~$\times$~4 Mueller matrix.\cite{Philipp_RSI} The MIR-OHE data is obtained at +6~T, 0~T, and -6~T, with the magnetic field parallel to the incoming infrared beam. Each surface cut is measured at one in-plane azimuth orientation. These measurements are performed at angle of incidence $\Phi_{a}$ = 45$^{\circ}$ and at temperature $T$ = 300~K. Additional measurements at zero field were performed at multiple in-plane orientations, and included into the data analysis. Note that the anisotropy of the effective mass parameter is determined by the anisotropy in the plasma frequency as discussed further below. The anisotropy of the plasma frequency parameter is determined at zero field and multiple azimuth orientations. Hence, OHE data were only performed at one azimuth orientation for each sample.   

In addition to the MIR-OHE measurements, zero magnetic field Mueller matrix data is measured at multiple azimuth orientations for each surface cut. The data is obtained using a commercially available MIR ellipsometer (IR-VASE, J.A. Woollam Co. Inc.) and the afore mentioned home-built ellipsometer in the spectral range of 150~cm$^{-1}$ to 1500~cm$^{-1}$ with a resolution of 2~cm$^{-1}$. The zero magnetic field data is not shown here, but is included in Ref.~\onlinecite{Schubert_2016}. These measurements are performed at $\Phi_{a}$ = 50$^{\circ}$, 60$^{\circ}$, and 70$^{\circ}$ and at room temperature. 

Ellipsometry is an indirect measurement technique which requires a physical parameterized model be fit to experimental data to determine the desired parameters.\cite{AzzamBook_1984} The model approach used here is very similar to that of Ref.~\onlinecite{Schubert_2016}. The two phase optical model consists of ambient air and $\beta$-Ga$_2$O$_3$ joined at a planar interface. The dielectric function tensor of $\beta$-Ga$_2$O$_3$ at long wavelengths consists of contributions from optical phonon modes and free charge carriers. These contributions are modeled using the eigendielectric displacement vector summation approach described in Refs.~\onlinecite{Schubert_2016,MockPRB2017CdWO4}. In this approach contributions from individual dielectric resonances, in this case phonon modes and free charge carriers, are added to a high frequency dielectric constant tensor $\boldsymbol{\varepsilon_{\infty}}$. The anharmonically broadened Lorentz oscillator model is used to represent phonon resonance contributions.\cite{MockPRB2017CdWO4} No substantial Drude contribution was detected in the off-diagonal components of the monoclinic dielectric function tensor.\cite{Schubert_2016} Thus, we employ an orthorhombic Drude model where three independent Drude contributions are added to the dielectric function response along axes \textbf{a} ($\varepsilon_{xx}$), \textbf{b} ($\varepsilon_{zz}$), and \textbf{c$^{\star}$} ($\varepsilon_{yy}$).

The magnetic field dependent free charge carrier contribution to the dielectric function tensor $\varepsilon^{\mathrm{FC}}(\omega)$ is described using the classical Drude formalism including the change induced by the Lorentz force\cite{Schubert_OHE,Philipp_RSI} 
\vspace{-0.1cm}
\begin{eqnarray}
\label{eqn:Equation1}
\varepsilon^{\mathrm{FC}} (\omega) = \frac{\boldsymbol{\omega_{\mathrm{p}}}^2}{
\begin{matrix} -\omega^2 \boldsymbol{I} - i\omega\boldsymbol{\gamma} + i\omega \begin{pmatrix} 0 & -b_\mathrm{z} & b_\mathrm{y}\\b_\mathrm{z} & 0 & -b_\mathrm{x}\\-b_\mathrm{y} & b_\mathrm{x} & 0 \end{pmatrix} \boldsymbol{\omega_{\mathrm{c}}} \end{matrix}}.
\end{eqnarray}
\noindent
Here, $\boldsymbol{I}$ is the identity matrix, and $\langle b_{\mathrm{x}}$, $b_{\mathrm{y}}$, $b_{\mathrm{z}} \rangle$ are the scalar components of magnetic field vector \textbf{B}, where each component is the projection along \textbf{x}, \textbf{y}, and \textbf{z}, respectively. At zero magnetic field, the classical Drude model parameters include the screened plasma frequency tensor $\boldsymbol{\omega_{\mathrm{p}}} = \sqrt{Nq^2/\varepsilon_0 \boldsymbol{\varepsilon_{\infty}} \boldsymbol{m^{\ast}}}$, and the plasma broadening tensor $\boldsymbol{\gamma} = q/ \boldsymbol{\mu m^{\ast}}$. These parameters depend on the free charge carrier properties which include effective mass $\boldsymbol{m^{\ast}}$, free charge carrier volume density $N$, and mobility $\boldsymbol{\mu}$, where $\boldsymbol{m^{\ast}}$ and $\boldsymbol{\mu}$ are diagonal second rank tensors. In the isotropic average approximation of a given tensor, its values are replaced by an isotropic scalar and the corresponding unit matrix. The parameter $\varepsilon_\mathrm{0}$ is the vacuum dielectric permittivity, and $q$ is the elementary electric charge. At non-zero magnetic field, the cyclotron frequency tensor is $\boldsymbol{\omega_{\mathrm{c}}} = q\rvert \boldsymbol{\mathrm{B}} \rvert/\boldsymbol{m^{\ast}}$.

\begin{figure}[ht]
\centering
\includegraphics[width=0.48\textwidth]{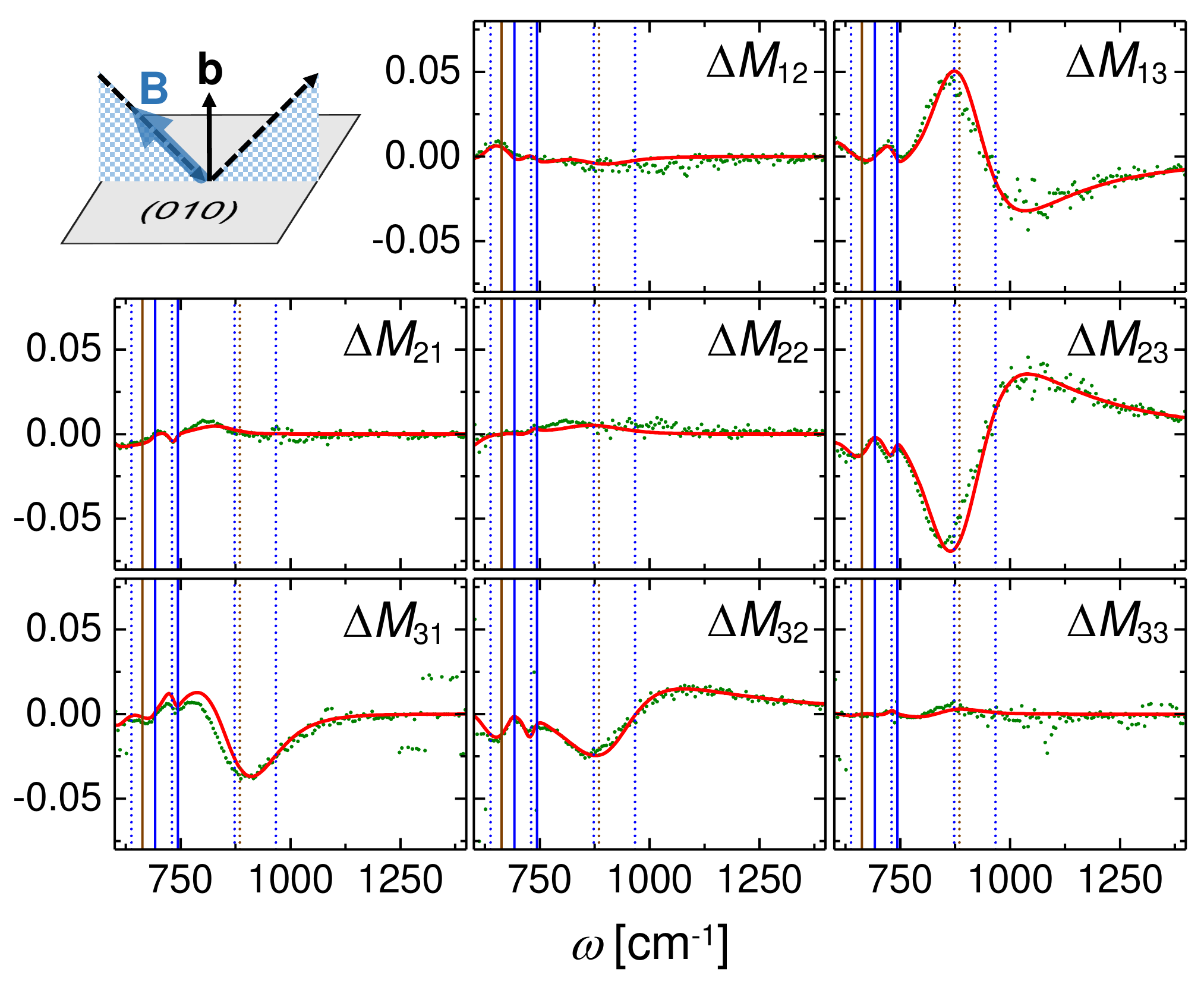}
\caption{
MIR-OHE experimental (green dots) and best-match model calculated (solid red lines) Mueller matrix difference spectra ($\Delta M_{\textrm{ij}} = M_{\textrm{ij}}(+6~\textrm{T}) - M_{\textrm{ij}}(-6~\textrm{T})$) for the ($010$) cut $\beta$-Ga$_2$O$_3$ sample at azimuth angle $\phi$~=~112.5$^{\circ}$. All measurements are performed at temperature $T$~=~300~K, and at angle of incidence $\Phi_{a}$~=~45$^{\circ}$. The magnetic field is parallel to the incoming infrared beam. Taken from Ref.~\onlinecite{Schubert_2016}, vertical lines signify the wave numbers of LPP (dotted lines) and transverse optical phonon modes (solid lines) polarized in the \textbf{a-c} plane (blue), and along the \textbf{b} axis (brown).
}
\label{Figure1}
\vspace{-0.5cm}
\end{figure}

\begin{figure}[ht]
\centering
\includegraphics[width=.48\textwidth]{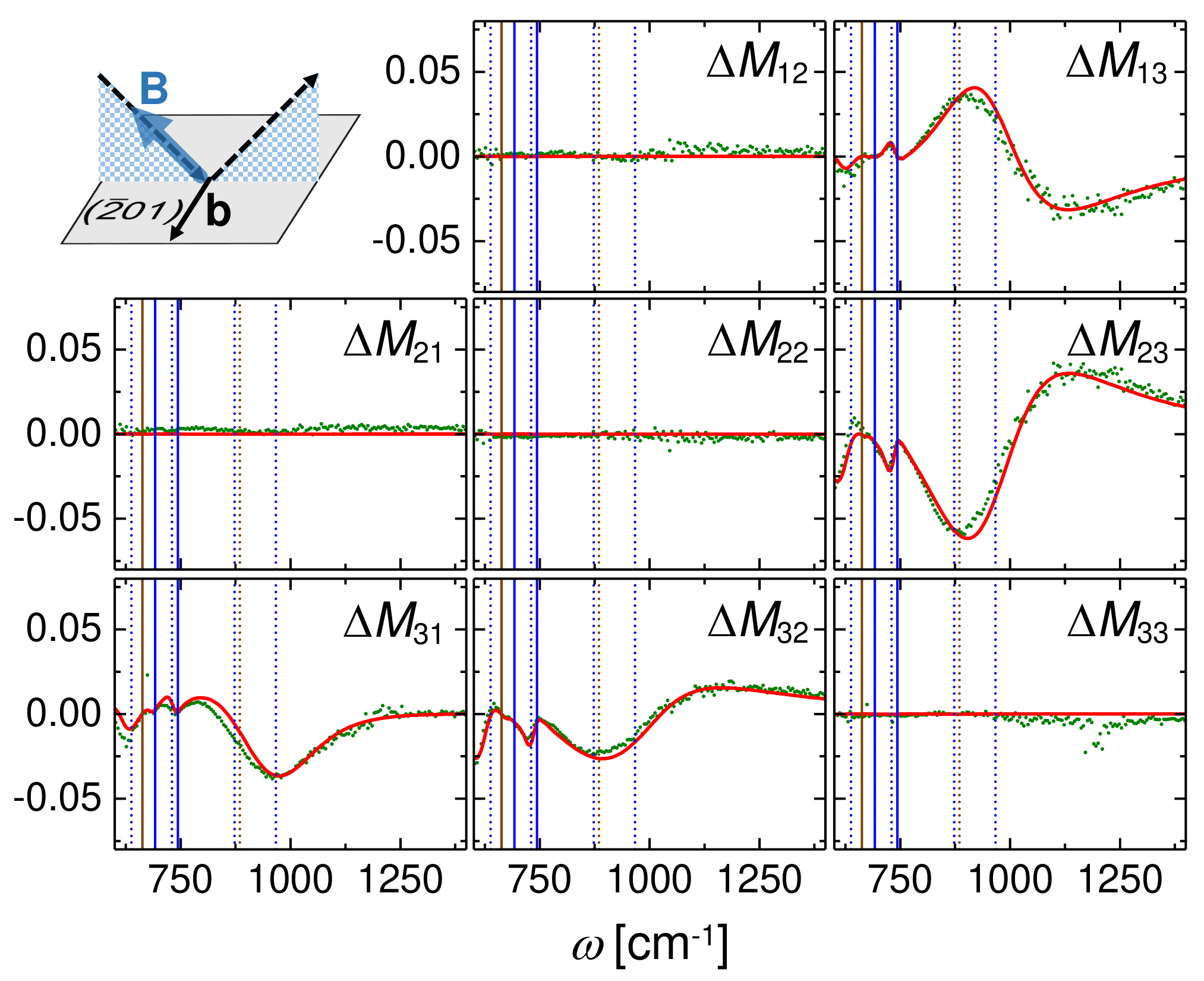}
\caption{Same as Fig.~\ref{Figure1} for the ($\bar{2}01$) cut $\beta$-Ga$_2$O$_3$ sample at azimuth angle $\phi$~=~181.7$^{\circ}$.}
\label{Figure2}
\vspace{-0.75cm}
\end{figure}

Figs.~\ref{Figure1} and ~\ref{Figure2} show experimental and best-match model calculated MIR-OHE difference spectra for the (010) and  ($\bar{2}01$) surface cut of $\beta$-Ga$_2$O$_3$, respectively. The MIR-OHE signals are strongest in the vicinity of the zero magnetic field reflectance minima for samples with a sufficient free charge carrier contributions.\cite{SchoecheAPL103_2013} For the samples investigated here, the reflectance minimum and MIR-OHE signal appear at the edge of the reststrahlen band at around 900~cm$^{-1}$. Due to the coupling of longitudinal optical phonon modes and free charge carriers, the so called longitudinal phonon plasmon (LPP) modes are now experimentally observed. Since the spectral locations of the reflectance minimas are governed by the LPP modes, the strongest MIR-OHE signatures occur in the vicinity of the highest frequency LPP modes, which are indicated by the vertical dotted lines in Fig.~\ref{Figure1} and Fig.~\ref{Figure2}.\cite{Schubert_2016} The unique shape of the signal is governed by changes in the dielectric function tensor due to phonon modes near this spectral range. We note that until this report, MIR-OHE difference data between positive and negative magnetic field were only seen in the off-block-diagonal Mueller matrix elements (i.e. $M_{13}$, $M_{23}$, $M_{31}$, $M_{32}$).\cite{SchoecheAPL103_2013} However, for the (010) surface (Fig.~\ref{Figure1}) a small difference is seen in the on-block-diagonal elements (i.e. $M_{12}$, $M_{21}$, $M_{22}$, $M_{33}$). This is due to the dielectric function tensor for the (010) surface at azimuth angle $\phi$~=~112.5$^{\circ}$ possessing non-zero off-diagonal elements at $\rvert \boldsymbol{\mathrm{B}} \rvert = 0$. In contrast, the ($\bar{2}01$) surface at azimuth angle $\phi$~=~181.7$^{\circ}$ possesses negligible off-diagonal tensor components at $\rvert \boldsymbol{\mathrm{B}} \rvert = 0$ since the \textbf{a}-\textbf{c} plane lies within the plane of incidence.

\begin{table}
\vspace{-0.25cm}
\caption{Results for isotropic average free charge carrier properties in $\beta$-Ga$_2$O$_3$. The theoretical Gau-PBE average effective mass parameters are calculated by taking the harmonic mean of the predicted values along the \textbf{a}, \textbf{b}, and \textbf{c} crystal axes, which can be found in Table~\ref{Table2}. Error bars shown correspond to the 90$\%$ confidence interval within the best-match model data analysis.}
\label{Table1}
\begin{ruledtabular}
\begin{tabular}{l l l}
Method  &  Parameter  &  Value\\
\noalign{\smallskip}\hline\noalign{\smallskip}
MIR-OHE for  &  $m^{\ast}$  &  $(0.284\pm0.013)m_{0}$\\
(010) surface$^\textrm{a}$  &  $N$  &  $(4.2\pm0.1)\times10^{18}~\text{cm}^{-3}$\\
         &  $\mu$  &  $(44\pm2)~\text{cm}^{2}/\text{Vs}$\\
\noalign{\smallskip}\hline\noalign{\smallskip}
MIR-OHE for  &  $m^{\ast}$  & $(0.283\pm0.011)m_{0}$\\
($\bar{2}01$) surface$^\textrm{a}$  &  $N$  &  $(5.9\pm0.1)\times10^{18}~\text{cm}^{-3}$\\
         &  $\mu$  &  $(43\pm1)~\text{cm}^{2}/\text{Vs}$\\
\noalign{\smallskip}\hline\noalign{\smallskip}
Gau-PBE$^\textrm{b}$  &  $m^{\ast}_{\textrm{avg}}$  &  $0.267m_{0}$\\
\end{tabular}
\end{ruledtabular}
\begin{flushleft}
\footnotesize{$^\textrm{a}${This work}}\\
\footnotesize{$^\textrm{b}${Theory, Ref.~\onlinecite{MockarXivGa2O3_2017}}}\\
\end{flushleft}
\vspace{-0.75cm}
\end{table}

Assuming separate sets of isotropic free charge carrier properties for the (010) and ($\bar{2}01$) surfaces, the model parameters are fit to the MIR-OHE difference data and zero magnetic field data simultaneously. The final best-match model fit is presented in Fig.~\ref{Figure1} and Fig.~\ref{Figure2}, and the resulting parameters are shown in Table~\ref{Table1}. The zero magnetic field data alone would allow one to determine the isotropic plasma frequency $\omega_{\mathrm{p}}$ and broadening $\gamma$, which are functions of $m^{\ast}$, $N$, and $\mu$. The addition of the MIR-OHE data in the analysis allows $m^{\ast}$, $N$, and $\mu$ to be accurately resolved. In order to improve the best match between model calculated and experimental data, the model for the (010) and ($\bar{2}01$) surfaces must be assigned independent sets of free charge carrier properties to account for a potentially different Sn dopant distribution and activation. However, each surface shares the same phonon mode parameters and $\boldsymbol{\varepsilon_{\infty}}$. Since the phonon mode parameters and $\boldsymbol{\varepsilon_{\infty}}$ in Ref.~\onlinecite{Schubert_2016} were derived assuming identical Drude parameters for the (010) and ($\bar{2}01$) surfaces, these quantities were also included in the best-match model fit to properly determine the free charge carrier properties. The analysis confirms the expected n-type conductivity for each surface cut. The azimuth angle $\phi$~=~112.5$^{\circ}$ and $\phi$~=~181.7$^{\circ}$ for the (010) and ($\bar{2}01$) surface, respectively, are determined by applying the zero magnetic field model to the zero-field MIR-OHE measurement. 

The electron effective mass parameters experimentally determined in this work are $m^{\ast} = (0.284\pm0.013)m_{0}$ for the (010) surface and $m^{\ast} = (0.283\pm0.011)m_{0}$ for the ($\bar{2}01$) surface. These fall within the broad range of values reported for various density functional theory calculations: (0.12 to 0.13)$m_{0}$,\cite{he2006} (0.22 to 0.30)$m_{0}$,\cite{MockarXivGa2O3_2017} (0.23 to 0.24)$m_{0}$,\cite{yamaguchi2004} (0.26 to 0.27)$m_{0}$,\cite{Furthmuller_2016} (0.27 to 0.28)$m_{0}$,\cite{varley2010,Peelaerspssb2015Ga2O3meff} (0.34)$m_{0}$,\cite{he2006_2} and (0.39)$m_{0}$.\cite{Ming-GangJMCA2014} The best-match model parameter results for the isotropically averaged mobility parameter $\mu$ for the two surfaces compare well with values determined previously by electrical Hall effect measurements for samples with similar free electron densities.\cite{SuzukiPSSC4_2007} Our best-match model parameter results for the electron density $N$ are in excellent agreement with the nominal density difference $N_{\mathrm{D}}-N_{\mathrm{A}} \approx 3\times 10^{18}$ cm$^{-3}$ provided by the crystal manufacturer.

Ueda~\textit{et al.} estimated electron effective mass parameters $m^{\ast}_{\textrm{b}}=0.5m_{0}~\textrm{to}~1.0m_{0}$ and $m^{\ast}_{\textrm{c}}=1.0m_{0}~\textrm{to}~2.0m_{0}$ from  optical transmission and electrical Hall effect measurements.\cite{UedaAPL71_1997} There is a rather large discrepancy between the effective mass parameters reported in this work and by Ueda \textit{et al}. A critical discussion of the results by Ueda~\textit{et al.} was given by Parisini~\textit{et al.} suggesting revision of data analysis in Ref.~\onlinecite{UedaAPL71_1997}.\cite{ParisiniSST31_2016}

\begin{table}
\vspace{-0.25cm}
\caption{Results for anisotropic free charge carrier properties in $\beta$-Ga$_2$O$_3$. Error bars shown correspond to the 90$\%$ confidence interval within the best-match model data analysis.}
\label{Table2}
\begin{ruledtabular}
\begin{tabular}{l l l}
Method  &  Parameter  &  Value\\
\noalign{\smallskip}\hline\noalign{\smallskip}
MIR-OHE for  &  $m^{\ast}_{\textrm{a}}$  &  $(0.288\pm0.044)m_{0}$\\
(010) surface$^\textrm{a}$   &  $m^{\ast}_{\textrm{b}}$  &  $(0.283\pm0.046)m_{0}$\\
         &  $m^{\ast}_{\textrm{c}^{\star}}$  &  $(0.286\pm0.044)m_{0}$\\
         &  $N$  &  $(4.1\pm0.3)\times10^{18}~\text{cm}^{-3}$\\
         &  $\mu_{\textrm{a}}$  &  $(45\pm4)~\text{cm}^{2}/\text{Vs}$\\
         &  $\mu_{\textrm{b}}$  &  $(42\pm4)~\text{cm}^{2}/\text{Vs}$\\
         &  $\mu_{\textrm{c}^{\star}}$  &  $(42\pm3)~\text{cm}^{2}/\text{Vs}$\\
\noalign{\smallskip}\hline\noalign{\smallskip}
MIR-OHE for  &  $m^{\ast}_{\textrm{a}}$  & $(0.295\pm0.039)m_{0}$\\
($\bar{2}01$) surface$^\textrm{a}$  &  $m^{\ast}_{\textrm{b}}$  &  $(0.276\pm0.037)m_{0}$\\
         &  $m^{\ast}_{\textrm{c}^{\star}}$  &  $(0.311\pm0.044)m_{0}$\\
         &  $N$  &  $(6.0\pm0.5)\times10^{18}~\text{cm}^{-3}$\\
         &  $\mu_{\textrm{a}}$  &  $(44\pm3)~\text{cm}^{2}/\text{Vs}$\\
         &  $\mu_{\textrm{b}}$  &  $(44\pm3)~\text{cm}^{2}/\text{Vs}$\\
         &  $\mu_{\textrm{c}^{\star}}$  &  $(41\pm3)~\text{cm}^{2}/\text{Vs}$\\
\noalign{\smallskip}\hline\noalign{\smallskip}
Gau-PBE$^\textrm{b}$  &  $m^{\ast}_{\textrm{a}}$  &  $0.224m_{0}$\\
         &  $m^{\ast}_{\textrm{b}}$	 &  $0.301m_{0}$\\
         &  $m^{\ast}_{\textrm{c}}$	 &  $0.291m_{0}$\\
\end{tabular}
\end{ruledtabular}
\begin{flushleft}
\footnotesize{$^\textrm{a}${This work}.}\\
\footnotesize{$^\textrm{b}${Theory, Ref.~\onlinecite{MockarXivGa2O3_2017}}.}
\end{flushleft}
\vspace{-0.75cm}
\end{table}

The anisotropy of $m^{\ast}$ may be defined by considering the ratios $(m^{\ast}_{\textrm{a}}/m^{\ast}_{\textrm{b}})$ and $(m^{\ast}_{\textrm{b}}/m^{\ast}_{\textrm{c}^{\star}})$. These quantities are comparable to the squares of the ratios of the plasma frequencies determined by the orthogonal Drude model approximation, via $(\omega_{\textrm{p},\textrm{b}}/\omega_{\textrm{p},\textrm{a}})^{2}=(m^{\ast}_{\textrm{a}}/m^{\ast}_{\textrm{b}})$ and $(\omega_{\textrm{p},\textrm{c}^{\star}}/\omega_{\textrm{p},\textrm{b}})^{2}=(m^{\ast}_{\textrm{b}}/m^{\ast}_{\textrm{c}^{\star}})$. The parameters $\omega_{\textrm{p},\textrm{a}}$, $\omega_{\textrm{p},\textrm{b}}$, and $\omega_{\textrm{p},\textrm{c}^{\star}}$ are the plasma frequencies corresponding to the \textbf{a}, \textbf{b}, and \textbf{$\textrm{c}^{\star}$} directions, respectively. Information about the plasma frequencies can be gathered without the use of magnetic fields. Generalized ellipsometry measurements at zero field and at multiple sample azimuth orientations, for both the (010) and ($\bar{2}01$) surfaces, were taken and subsequently analyzed simultaneously with the MIR-OHE data. This approach provided sufficient sensitivity to determine the anisotropy of the free charge carrier parameters. The resulting effective mass parameters are shown in Table~\ref{Table2}. The ratios are $(m^{\ast}_{\textrm{a}}/m^{\ast}_{\textrm{b}})=(1.02^{+0.38}_{-0.28})$ and $(m^{\ast}_{\textrm{b}}/m^{\ast}_{\textrm{c}^{\star}})=(0.99^{+0.37}_{-0.27})$ for the (010) surface, where the upper and lower scripted numbers refer to the upper and lower uncertainty limit of the effective mass parameter ratio. The upper/lower limits come from taking the ratios within the maximum/minimum parameter deviations in the numerator and denominator using the mass parameters and error bars as shown in Table~\ref{Table2}. For the ($\bar{2}01$) surface, the ratios are $(m^{\ast}_{\textrm{a}}/m^{\ast}_{\textrm{b}})=(1.07^{+0.33}_{-0.25})$ and $(m^{\ast}_{\textrm{b}}/m^{\ast}_{\textrm{c}^{\star}})=(0.89^{+0.29}_{-0.22})$. Our findings suggest as small deviation from isotropy, however, which is well within the uncertainty limits for both surfaces investigated. Nonetheless, the possibility of a small anisotropy would be consistent with recent theoretical investigations.\cite{yamaguchi2004,he2006,Peelaerspssb2015Ga2O3meff,Furthmuller_2016} Yamaguchi calculated the electron effective mass ratios of $(m^{\ast}_{\textrm{a}'}/m^{\ast}_{\textrm{b}'})=0.96$ and $(m^{\ast}_{\textrm{b}'}/m^{\ast}_{\textrm{c}'})=1.07$ at the $\Gamma$ point, where $m^{\ast}_{\textrm{a}'}$, $m^{\ast}_{\textrm{b}'}$, and $m^{\ast}_{\textrm{c}'}$ are diagonal Cartesian effective mass tensor components.\cite{yamaguchi2004} Furthm\"{u}ller and Bechstedt predict ratios of $(m^{\ast}_{\textrm{a}'}/m^{\ast}_{\textrm{b}'})=1.01$ and $(m^{\ast}_{\textrm{b}'}/m^{\ast}_{\textrm{c}'})=1.03$.\cite{Furthmuller_2016} He \textit{et al.} finds ratios of $(m^{\ast}_{\textrm{a}^{\star}}/m^{\ast}_{\textrm{b}^{\star}})=0.95$ and $(m^{\ast}_{\textrm{b}^{\star}}/m^{\ast}_{\textrm{c}^{\star}})=1.05$.\cite{he2006} Recent density functional calculations using the Gau-PBE approach predict ratios of $(m^{\ast}_{\textrm{a}}/m^{\ast}_{\textrm{b}})=0.74$ and $(m^{\ast}_{\textrm{b}}/m^{\ast}_{\textrm{c}})=1.03$ for undoped $\beta$-Ga$_2$O$_3$.\cite{MockarXivGa2O3_2017} We note that the theoretical results reported so far are inconsistent, however, all theoretical predicted ratios could fall within our experimental error bars and no conclusive statement about a finite anisotropy of the effective electron mass parameter can be made at this point. Within our uncertainty limits, the mobility parameter is found to be essentially isotropic. This is consistent with previous theoretical investigations for intrinsic mobility,\cite{KangJPCM29_2017} experimental Hall effect measurements using the bar method\cite{VilloraJCG270_2004} and using the Van der Pauw method.\cite{GolzIWGO_2017,IrmscherJAP110_2011} A nearly isotropic mobility is also reported for electron channel mobility in silicon-doped Ga$_2$O$_3$ metal-oxide-semiconductor field-effect transistors (MOSFETs).\cite{WongJJAP55_2016}

This work was supported by the Swedish Research Council (VR) under Grant No. 2013-5580 and 2016-00889,
the Swedish Governmental Agency for Innovation Systems (VINNOVA) under the VINNMER international qualification
program, Grant No. 2011-03486, the Swedish Government Strategic Research Area in Materials Science on Functional
Materials at Link\"oping University, Faculty Grant SFO Mat LiU No 2009 00971, and the Swedish Foundation for Strategic
Research (SSF), under Grant Nos. FL12-0181 and RIF14-055. The authors further acknowledge financial support by the University of Nebraska-Lincoln, the J.~A.~Woollam Co., Inc., the J.~A.~Woollam Foundation and the National Science Foundation (awards MRSEC DMR 1420645, CMMI 1337856 and EAR 1521428).


\bibliography{Ga2O3-OHE_APL_2017_bib_file}
\end{document}